\documentclass[preprint,12pt]{elsarticle}
\usepackage{graphicx}
\usepackage{amssymb}
\usepackage{bm}
\usepackage{subfigure}
\graphicspath {{./}}
\journal{J. Nucl. Mater.}

\begin{document}

\begin{frontmatter}

  \title{Structure, elastic and bonding properties of hcp Zr$_{x}$Ti$_{1-x}$ binary alloy from first-principles calculations                              }

\author{S. J. Hou$^1$, H. P. Lei$^1$, S. C. Huang$^1$, Z. Zeng$^{1,2}$*}

\address{*Corresponding Author: zzeng@theory.issp.ac.cn \\
$^1$Key Laboratory for Materials Physics, Institute of Solid State
Physics, \\ Chinese Academy of Sciences, Hefei, 230031, China \\
$^2$Department of Physics, University of Science and Technology of China, \\
Hefei, 230026, China \\
}
\begin{abstract}
First principles calculations were performed to study the structural, 
elastic, and bonding properties 
of hcp Zr$_{x}$Ti$_{1-x}$ binary alloy. The special quasi-random structure (SQS) method is employed to
mimic the random hcp Zr$_{x}$Ti$_{1-x}$ alloy.  It is found that Bulk modulus, B, Young's 
modulus, E, and shear modulus, G, exhibit decreasing trends as increasing
the amount of Zr. A ductile behavior Zr$_{x}$Ti$_{1-x}$ is predicted 
in the whole composition range. 
In terms of Mulliken charge analyze, it is found that the element Ti behave much more electronegative
than Zr in hcp Zr$_{x}$Ti$_{1-x}$ alloy, and the amount of charge transfer between them is
approximately linear to the number of Ti(Zr) surrounding Zr(Ti).
\end{abstract}


\end{frontmatter}
\section{Introduction}
Group IV transition metals including titanium (Ti), zirconium (Zr), and hafnium (Hf) and their alloys
have attracted enormous research and technological interests due to their excellent properties 
such as high strength-to-weight, high rigidity-to-weight ratio, low thermal neutron absorption cross
section and good corrosion resistance \cite{Zhilyaev2009}. 
Their narrow d-band characterized as in the midst of a broad sp-band is the origin of 
scientific interest. And the pressure-induced electrons transfer from sp-band to 
d-band is the driving force behind the 
structural and electronic transitions.
As a member of the group IV alloys, Zr-Ti alloys have wide
 applications in aerospace, medical, nuclear industries. 

 At ambient condition, pure zirconium and titanium both crystallizes in hexagonal closed 
 packed(hcp) structure ($\alpha$ phase). For zirconium, experiments\cite{HuiXia1990,HuiXia1991} 
 found that it undergoes
 a crystallographic phase transition from hcp($\alpha$ phase) to another hexagonal structure
 ($\omega$ phase) at pressure of 2-7 GPa, and it will transform to 
 the body-centered-cubic structure ($\beta$ phase) at the pressure of 30-35 GPa.
 However, for Ti, the experimental phase transition order at room temperature is alpha omega
 gama delta, its beta phase has not been found until 216 GPa\cite{Y.Akahama2001}.
For the ZrTi system, experimentally,  it is characterized 
by full solubility of its components\cite{I.O.Bashkin2000,I.O.Bashkin2003}. 
As in pure zirconium and titanium, three phases($\alpha$, $\beta$ and $\omega$) are 
observed in the  ZrTi system. At ambient condition, they crystallize in hexagonal close-packed (hcp) 
structure
($\alpha$ phase), and transforms to body-centered cubic (bcc) $\beta$ phase under high temperature and a
three atoms hexagonal structure ($\omega$ phase) under pressure.
  The aim of the present paper is to use first-principles 
  calculations to theoretically investigate the compositional  
  dependence of the structural, elastic and bond properties of 
  the Zr$_{x}$Ti$_{1-x}$ binary alloy system.

\section{Methods}
In present work,the first-principles 
DFT calculations are performed using the projector augmented wave (PAW) \cite{Blochl1994} as implemented
in the Vienna ab initio simulation package (VASP) \cite{Kresse1996}.
To describe the exchange-correlation potential, the Perdew-Burke-Ernzerhof (PBE) \cite{Perdew1996} form 
of the generalized gradient approximation (GGA) is employed.
The Zr 4d$^2$5s$^2$5p$^0$ and the Ti 4d$^2$5s$^2$5p$^0$ orbitals are treated as valance
electrons. 
To get accurate results, the plane wave cut-off energy is chosen as 400 eV. The Brillouin-zone 
integrations are performed using $\gamma$-centered grids of kpoints of 
25 $\times$ 25 $\times$ 25 for $\alpha$-Zr, 
25 $\times$ 25 $\times$ 25 for $\alpha$-Ti, 15$\times$ 15 $\times$ 15 for
Zr$_{x}$Ti$_{1-x}$ in terms of Monkhorst-Pack scheme \cite{Monkhorst1976}. 
The geometries are optimized until the Hellmann-Feynman forces
are less than 0.01 eV/{\AA}, and the total energy is relaxed until the difference value becoming
smaller than 10-5 eV.

Using the wavefunction  obtained from the DFT calculation, 
the QUAMBO method \cite{Lu2004,Chan2007,Qian2008,Yao2010}
was implemented to exactly down-fold the occupied states to a representation
of a minimal-basis set without losing any electronic structure information.
The constructed orbitals are atomiclike and highly localized, and they are adapted to perform the 
chemical bonding analysis to the interaction mechanism of Ti and Zr.

The concept of SQS was first proposed  by Zunger et al. \cite{Zunger1990} to 
mimic disordered (random) solution. 
There exists a one-to-one correspondence between a given structure and a set of correlation functions,
which is the key for SQS methods. In a substitutional binary alloy case, the correlation function 
$\prod_{k,l}$ for a figure (cluster) $f(k,l)$ with $k$ vertices and separated by an $l$th neighbor 
distance is defined as follows: 
$$\prod_{k,l}=\frac{1}{N_{k,l}}\sum_{k,l}\sigma_1\sigma_2\cdots\sigma_k \eqno(1)$$ where $\sigma_k$ is a 
spinlike variable which takes the value of +1 or -1 depending on whether the atomic site is 
occupied by an $A$ or $B$ atom. Specially, for a random alloy of $A_{1-x}B_x$, Eq. (1) is simply by
${(2x-1)}^k$. The optimum SQS for a given number of atoms is the one that best matches with
the correlation function of the random alloy. 
In the present work, SQS models were generated using the Monte Carlo algorithm 
implemented by Walle et al.\cite{VandeWalle2013}.  
Their pair correction functions $\Pi_{2,l}$(l up to the 11th nearest neighbor) are shown in Table 1. For
Zr$_{4}$Ti$_{12}$ ,the $\Pi_{2,l}$ match the random ones well until l=6; and l=8 for Zr$_{8}$Ti$_{8}$.
An example(Zr$_{4}$Ti$_{12}$) of SQSs generated in present work is given in Fig. 1.

In principle, the point group symmetry of the original alloy is broken by the SQS method. As a result, 
there will be 21 elastic constant elements for a SQS model \cite{Tasnadi2012a}. 
Traditional, the energy-strain approach \cite{LePage2001}
and the stress-strain approach \cite{LePage2002a} are two ways of 
calculating single crystal elastic constants from 
first-principles calculations. In order to obtain 21 elastic constant with the energy-strain approach, 
we need to impose 21 independent deformation on the original structure. Extremely computing power
is required.  
In order to calculate single crystal elastic constants from
first-principles calculations,the stress-strain approach was adopted in the present work.
A set of small strains $\bm\varepsilon=(\varepsilon_1\ \varepsilon_2\ \varepsilon_3\ \varepsilon_4\ 
\varepsilon_5\ \varepsilon_6)$ (where $\varepsilon_1$, $\varepsilon_2$, and $\varepsilon_3$ are the 
normal strains,
$\varepsilon_4$, $\varepsilon_5$, and $\varepsilon_6$ are the shear 
strains in Voigt's notation) is imposed on a crystal, the deformed structure lattice vectors ($\overline
{\bm{R}}$) are obtained by transforming the original one ($\bm{R}$) as follows:

\begin{equation}       
  \overline{\bm{R}}=\bm{R}
\left(                 
\begin{array}{ccc}    
  1+\varepsilon_1 & \varepsilon_6/2 & \varepsilon_5/2\\  
\varepsilon_6/2 & 1+\varepsilon_2 & \varepsilon_4/2\\  
\varepsilon_5/2 & \varepsilon_4/2 & 1+\varepsilon_3\\  
\end{array}
\right)                
\end{equation}
As a result, a set of stress $\bm{t}=(t_1\ t_2\ t_3\ t_4\ t_5\ t_6)$ is determined by first-principles
calculations in this work. In present work, we apply the following six linearly independent sets of 
strains \cite{Shang2007}

\begin{equation}       
\left(                 
\begin{array}{cccccc}    
x&0&0&0&0&0\\
0&x&0&0&0&0\\
0&0&x&0&0&0\\
0&0&0&x&0&0\\
0&0&0&0&x&0\\
0&0&0&0&0&x\\
\end{array}
\right)                
\end{equation}
with $x=\pm{0.007}$,
Using a $6\times6$ elastic constants matrix, $\bm{C}$, with components of $C_{ij}$ in Voigt's notation, 
the generalized Hooke's law is expressed as $\bm{t}=\bm{\varepsilon}\bm{C}$. Consequently, the stiffness
constants matrix is obtained from $$\bm{C}={\bm{\varepsilon}}^{-1}\bm{t},$$ where ``-1'' represents 
the pseudo-inverse, which can be solved by the singular value decomposition method. Finally, we get the
macroscopic hcp elastic constant, $\overline{C}_{11},\overline  {C}_{12},
\overline{C}_{13},\overline  {C}_{33},\overline  {C}_{44}$, by averaging \cite{Moakher2006}

\begin{equation}
\begin{array}{l}
  \overline  {C}_{11}  = 3(C_{11} + {C_{22}})/8+C_{12}/4 + {C_{66}}/2 \\ 
  \overline { C}_{12}  = ({C_{11}} + {C_{22}})/8+ 3{C_{12}}/4 - {C_{66}}/2 \\ 
  \overline { C}_{13}  = ({C_{13}} + {C_{23}})/2 \\ 
  \overline { C}_{33}  = {C_{33}} \\ 
  \overline { C}_{44}  = ({C_{44}} + {C_{55}})/2 \\ 
 \end{array}
\end{equation}

\section{Results and discussion}

\subsection{Elastic properties        }
Elastic constants  are very important because they can measure the resistance and mechanical
properties of a solid to external stress or pressure. All independent elastic
constants of Zr$_{x}$Ti$_{1-x}$ (x=0,0.25,0.5,0.75,1) are calculated using strain-stress method in
present work, and the results are summarized in Table 2. 
Small deviations from a perfect hcp structure are observed in the elastic tensors of the SQS models.
These elastic constants decrease as Zr content increases except $C_{12},C_{13},C_{23}$.
According to Eq.(3), the averaged $C_{11},C_{12},C_{13},C_{33},C_{44}$ are obtained 
for hcp crystals. The obtained constants of all composition
meet the requirement of the Born stability criteria\cite{Nye1985} for hcp system
$$C_{11}>0, C_{44}>0, C_{11}>|C_{12}|,(C_{11}+2C_{12})C_{33}>2{C^2_{13}}.$$,
The polycrystalline bulk modulus B, shear modulus G are deduced from the Voigt-Reuss-Hill(VRH) 
approach \cite{Hill1952}. Young's modulus and Poisson's ($\upsilon$) ratio are calculated by the following
formulas:
$$E = 9BG/(3B+G), \upsilon = (3B-2G)/[2(3B+G)]$$. The results and B/G are listed in Table 3.
The bulk moduli 
for all the composition show a excellent agreement with those obtained by fitting to a
Birch-Murnaghan equation of state (list in Table 1), which is a proof of consistency and reliability
of our calculations.
Additionally, the deduced bulk moduli B  change smoothly and decrease with increasing the amount 
of Zr, while the Young's modulus E and shear modulus G show the same trend until x=0.75.

Empirical, there are two common ways to judge a material ductile or brittle.
According Pugh's suggestion, a higher ratio( $>$ 1.75) of bulk to shear 
moduli, B/G, indicates ductile behavior
\cite{Pugh1954}.  
Another is the Poisson's ration, the transsion from brittleness to ductility occurs 
when $\upsilon$$\approx$1/3\cite{Frantsevich1983}.
Poisson's ratio, $\upsilon$, and the B/G ratio as a function of Zr content, x, are listed in Table 3.
Both criterion confirms the ductile behavior of Zr$_{x}$Ti$_{1-x}$ over the whole composition range.
The value of B/G and $\upsilon$ for ZrTi alloy are higher than the one for pure metal, indicating 
the ductility of Zr is enhanced when alloying with Ti.

\subsection{Mulliken charge}
In order to understand the bond property between atoms,
the atomic Mulliken charge of Zr-Ti binary alloy are investigated, and the results are 
given in Table 4,  
which clearly indicates that Ti atoms gain electrons while Zr atoms loss electrons .
 To investigate the origin of charge transfer, 
 we further investigate the relationship between charge transfer
and the amount of other element atom of its nearest neighbors 
by fitting the data using a line relationship.
The results are showed in Fig. 2. Obviously, they are line-like. So we conclude that 
the charge transfer is mainly determined by the number of other element atom in its nearest 
neighbors.

\section{Conclusion}
The structural, elastic and bond properties of  the Zr$_{x}$Ti$_{1-x}$ alloy have been studied using 
first-principles calculations. The SQS method are adopted to mimic the ZrTi random system.
It can be found that hcp structured Zr$_{x}$Ti$_{1-x}$ is a ductile material over the whole 
composition. The bulk Bulk modulus, B, Young's 
modulus, E, and shear modulus, G all have a decreasing trend with increasing the content of Zr.
The effect of alloy will enhanced the ductility of pure metal Zr or Ti.
From mulliken charge analysis, we could conclude that the amount of charge transfer is determined by
the number of other element in its nearest neighbors, and there is a line-like relationship 
between them.

\newpage
\noindent \textbf{Acknowledgments}\\

\noindent This work was supported by the National Science Foundation of China under Grant Nos. 11275229
\& NSAF U1230202, special Funds for Major State Basic Research Project of China (973) under Grant
No. 2012CB933702, Hefei Center for Physical Science and Technology under Grant No. 2012FXZY004,
Anhui Provincial Natural Science Foundation under Grant No. 1208085QA05, and Director Grants of
CASHIPS. Part of the calculations were performed at the Center for Computational Science of CASHIPS,
the ScGrid of Supercomputing Center, and the Computer Network Information Center of the Chinese Academy of
Sciences.

\bibliographystyle{elsarticle-num}
\bibliography{Refs}
\newpage

\begin{figure}[!htbp]
\centering
\includegraphics[width=0.8\textwidth]{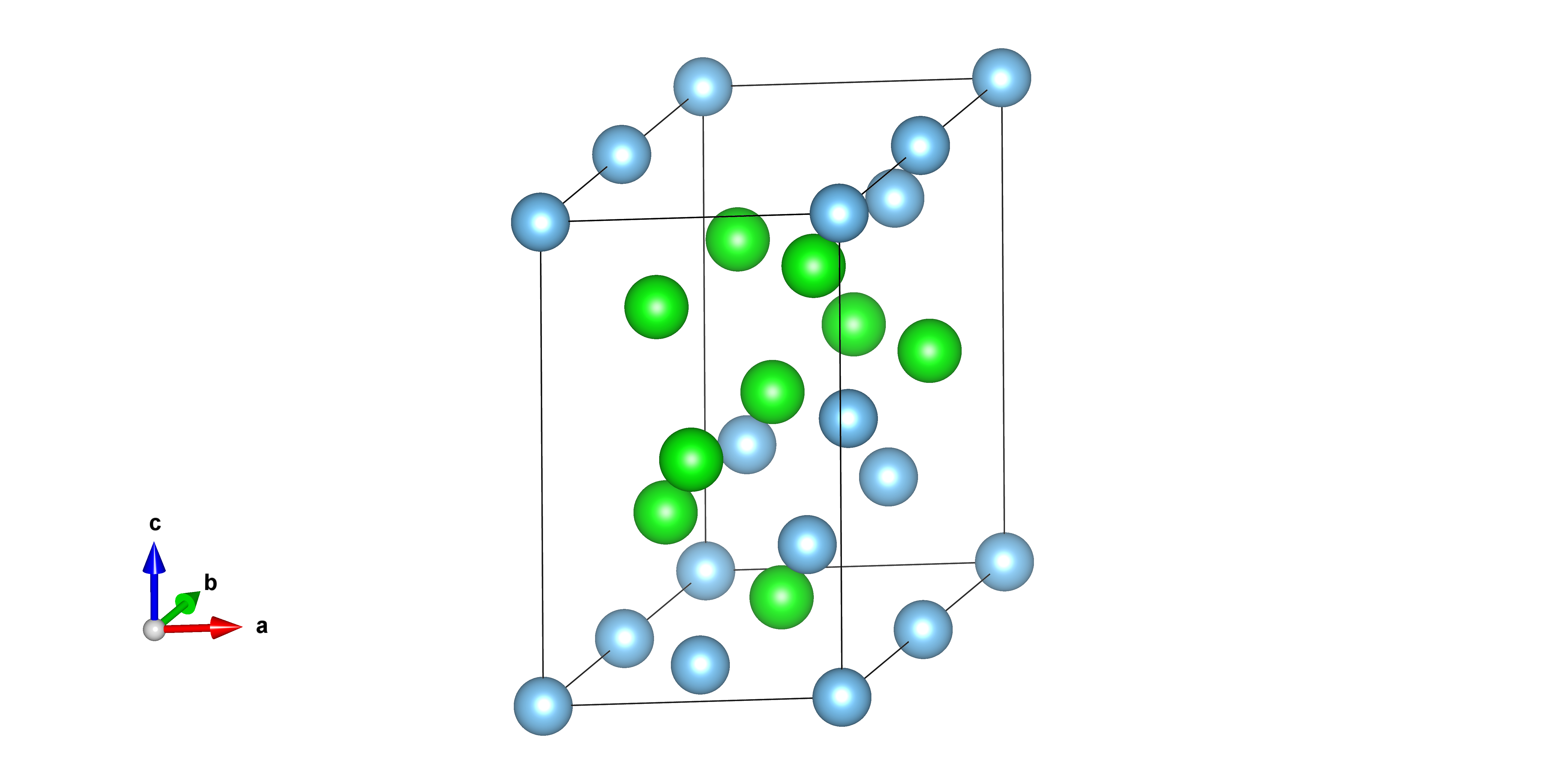}
\caption{The SQS structure of Zr$_{x}$Ti$_{1-x}$ with x = 0.5 represented by VESTA\cite{Momma2011}.
Zr: green spheres; Ti: blue spheres}
\label{structure}
\end{figure}

\begin{table*}[htbp]
\scriptsize
\caption{Pair correction functions $\Pi_{2,l}$(l up to the 11th nearest neighbor) for the random structures and SQS structures
}
\begin{center}
  \resizebox{\textwidth}{!}{
  \begin{tabular}{cccccccccccc} 
  \hline
Structure&$\Pi_{2,1}$&$\Pi_{2,2}$&$\Pi_{2,3}$&$\Pi_{2,4}$&$\Pi_{2,5}$&$\Pi_{2,6}$&$\Pi_{2,7}$&$\Pi_{2,8}$&$\Pi_{2,9}$&$\Pi_{2,10}$&$\Pi_{2,11}$\\ \hline
Random(Zr$_{0.5}$Ti$_{0.5}$)&0&0&0&0&0&0&0&0&0&0&0\\
SQS(Zr$_{8}$Ti$_{8}$)&0&0&0&0&0&0&0&-0.333333&0&0&0\\
Random(Zr$_{0.25}$Ti$_{0.75}$)&0.25&0.25&0.25&0.25&0.25&0.25&0.25&0.25&0.25&0.25&0.25\\
SQS(Zr$_{4}$Ti$_{12}$)&0.25&0.25&0.25&0.25&0.25&0.333333&0.458333&0.166667&0.25&0.166667&0.25\\
\hline
\end{tabular}
}
\end{center}
\label{Table.5}
\end{table*}

\begin{table*}[htbp]
\scriptsize
\caption{The elastic constants (in GPa) calculated by stress-strain approach
}
\begin{center}
  \resizebox{\textwidth}{!}{
  \begin{tabular}{cccccccccccccccccccccc} 
  \hline
Zr$_{x}$Ti$_{1-x}$&$C_{11}$&$C_{22}$&$C_{33}$&$C_{12}$&$C_{13}$&$C_{23}$&
$C_{44}$&$C_{55}$&$C_{66}$&$C_{14    }$&$C_{15}$&
$C_{16}$&$C_{24}$&$C_{25}$&$C_{26}$&$C_{34}$&$C_{35}$&$C_{36}$&$C_{45}$&$C_{46}$&$C_{56}$\\
 \hline
0.00&176.7&177.7&188.3&89.5&82.9&84.0&40.6&40.3&52.6&0.0&0.0&0.0&0.0&0.0&0.0&0.0&0.0&0.0&0.0&0.0&0.0\\
0.25&163.2&153.1&147.5&76.1&89.6&87.2&40.7&32.3&42.3&2.8&-1.3&5.2&-7.0&-3.2&0.7&5.5&6.1&3.0&-2.3&2.2&1.8\\
0.50&148.7&135.6&134.6&83.8&79.2&88.6&31.5&36.1&28.8
&-2.7&7.0&0.7&-0.9&0.8&-6.1&1.3&1.5&4.9&-1.0&1.1&-2.9\\
0.75&145.8&129.4&129.5&75.4&84.4&82.7&28.9&23.6&33.8&3.1&3.9&8.4&-2.8&-0.7&1.6&4.1&0.8&-0.9&0.8&4.2&0.2\\
1.00&153.1&153.0&163.1&63.6&70.5&70.7&26.2&26.2&45.2&0.0&0.0&0.0&0.0&0.0&0.0&0.0&0.0&-0.1&0.1&0.0&0.0\\
\hline
\end{tabular}
}
\end{center}
\label{Table.1}
\end{table*}

\begin{table*}[htbp]
\scriptsize
\caption{The elastic constants C$_{ij}$(GPa), shear modulus G(GPa), bulk modulus B(GPa),
Young's modulus E(GPa), Poission ratio $\upsilon$ at various composition for ZrTi alloy
}
\begin{center}
  \resizebox{\textwidth}{!}{
  \begin{tabular}{ccccccccccc} 
  \hline
Zr$_{x}$Ti$_{1-x}$&$\overline{C}_{11}(GPa)$&$\overline{C}_{12}(GPa)$&$\overline{C}_{13}(GPa)$&$\overline{C}_{33}(GPa)$&$\overline{C}_{44}(GPa)$&B(GPa)&G(GPa)&B/G&E(GPa)&$\upsilon$\\ \hline
0.00&181.6&85.1&83.5&188.3&40.5&117.3&45.6&2.6&121.0&0.33\\
0.25&158.8&75.5&88.4&147.5&36.5&107.7&36.8&2.9&99.1&0.35 \\
0.50&142.0&84.0&83.9&134.6&33.8&102.4&30.3&3.4&82.7&0.37 \\
0.75&138.9&74.0&83.6&129.5&26.3&98.9&27.8&3.5&76.3&0.37  \\
1.00&153.3&63.4&70.6&163.1&26.2&97.5&36.0&2.7&96.0&0.34  \\
\hline
\end{tabular}
}
\end{center}
\label{Table.2}
\end{table*}

\begin{table*}[htbp]
\scriptsize
\caption{Charge transfer according to Mulliken charge                       
}
\begin{center}
  \resizebox{\textwidth}{!}{
  \begin{tabular}{ccccccccccc} 
  \hline
Zr$_{0.25}$Ti$_{0.75}$&Charge Transfer&Zr$_{0.5}$Ti$_{0.5}$&Charge Transfer&Zr$_{0.75}$Ti$_{0.25}$&Charge Transfer&\\ \hline
Zr1  &-0.2000  &Zr1  &-0.1509  &Zr1  &-0.1382 \\                                                       
Zr2  &-0.4074  &Zr2  &-0.1653  &Zr2  &-0.0311 \\
Zr3  &-0.2229  &Zr3  &-0.2061  &Zr3  &-0.1335 \\
Zr4  &-0.5269  &Zr4  &-0.0621  &Zr4  &-0.2778 \\
Ti5  &0.0943   &Zr5  &-0.1748  &Zr5  &0.0042  \\
Ti6  &0.0147   &Zr6  &-0.3780  &Zr6  &-0.0871 \\
Ti7  &0.1865   &Zr7  &-0.2211  &Zr7  &-0.1850 \\
Ti8  &0.2337   &Zr8  &-0.3854  &Zr8  &-0.0124 \\
Ti9  &-0.0051  &Ti9  &0.1597   &Zr9  &-0.2204 \\
Ti10 &0.0903   &Ti10 &0.0870   &Zr10 &-0.2293 \\
Ti11 &0.1793   &Ti11 &0.1600   &Zr11 &-0.0562 \\
Ti12 &-0.0097  &Ti12 &0.2633   &Zr12 &-0.0640 \\
Ti13 &0.2233   &Ti13 &0.4042   &Ti13 &0.3147  \\
Ti14 &0.2843   &Ti14 &0.1512   &Ti14 &0.3467  \\
Ti15 &0.0080   &Ti15 &0.3217   &Ti15 &0.3187  \\
Ti16 &0.0577   &Ti16 &0.1969   &Ti16 &0.4507  \\
\hline
\end{tabular}
}
\end{center}
\label{Table.2}
\end{table*}

\begin{figure}[!htbp]
\centering
\subfigure[\($Zr$_{0.25}$Ti$_{0.75}\)]{
\includegraphics[width=0.4\textwidth]{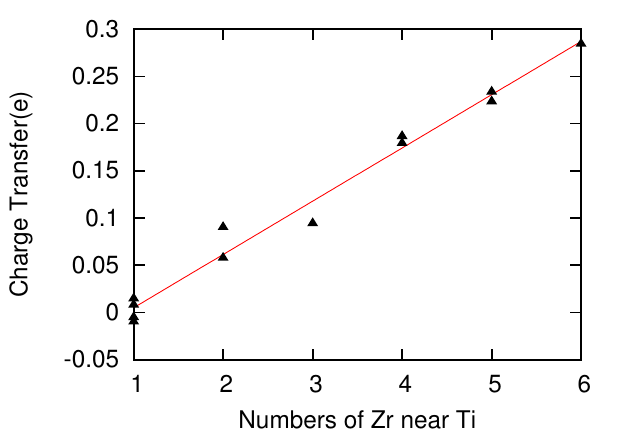}}
\subfigure[\($Zr$_{0.25}$Ti$_{0.75}\)]{
\includegraphics[width=0.4\textwidth]{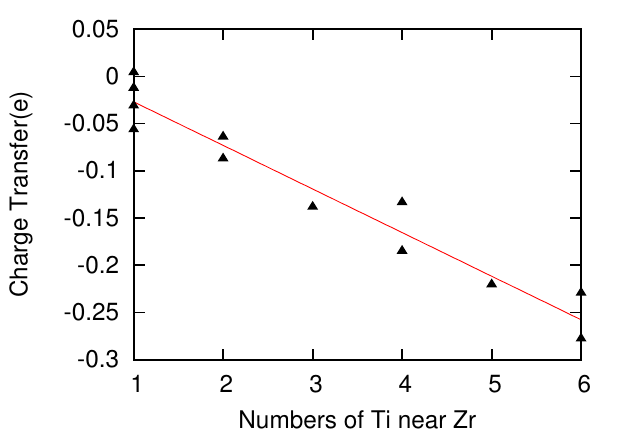}}
\\
\subfigure[\($Zr$_{0.5}$Ti$_{0.5}\)]{
\includegraphics[width=0.4\textwidth]{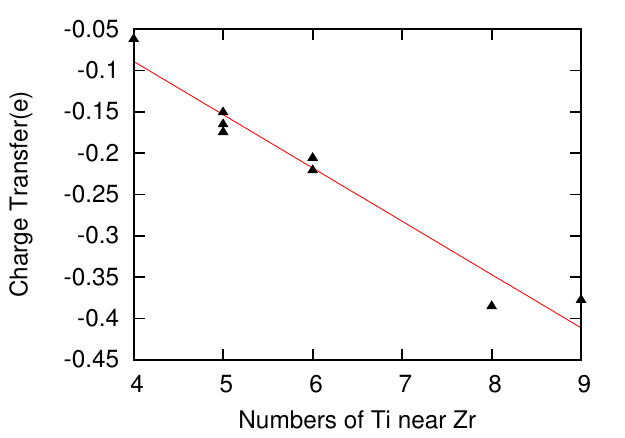}}
\subfigure[\($Zr$_{0.5}$Ti$_{0.5}\)]{
\includegraphics[width=0.4\textwidth]{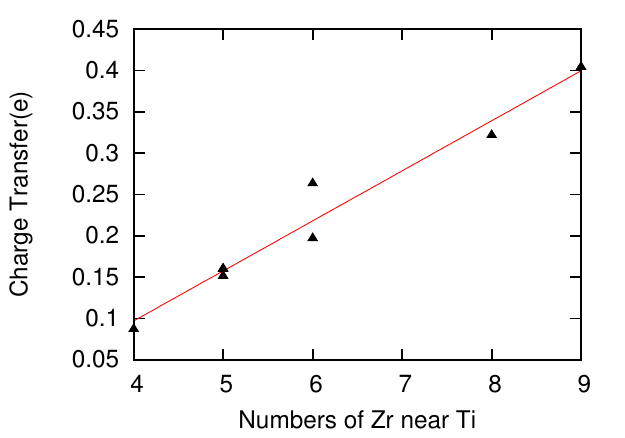}}
\caption{The charge transfer of ZrTi binary alloy with the mulliken charge method}
\label{charge transfer}
\end{figure}

\end{document}